\documentclass[preprint,proceedings]{rmaa}

\usepackage{paralist}

\usepackage{psfrag,color}

\SetYear{2005}
\SetConfTitle{Triggering Relativistic Jets}

\title{Present and Future Millimeter VLBI Imaging of Jets in AGN:

      The Case of NRAO\,150}

\author{
  I. Agudo,\altaffilmark{1}
  T.P. Krichbaum,\altaffilmark{1}
  U. Bach,\altaffilmark{1,2}
  A. Pagels,\altaffilmark{1}
  D. Graham,\altaffilmark{1}
  W. Alef,\altaffilmark{1}
  A. Witzel,\altaffilmark{1}
  J.A. Zensus,\altaffilmark{1}
  M. Bremer,\altaffilmark{3}
  M. Grewing\altaffilmark{3}
  and  H. Ter\"asranta\altaffilmark{4}}

\altaffiltext{1}{Max-Planck-Institut f\"ur Radioastronomie, Bonn, Germany.}
\altaffiltext{2}{Osservatorio Astronomico di Torino, Pino Torinese, Italy.}
\altaffiltext{3}{Institut de Radio Astronomie Millim\'etrique, Grenoble, France.}
\altaffiltext{4}{Mets\"ahovi radio observatory, Kylm\"al\"a, Finland.}

\shortauthor{Agudo et al.}
\shorttitle{Millimeter VLBI Imaging of Jets in AGN:
           The Case of NRAO\,150}

\fulladdresses{
\item Iv\'an Agudo, Thomas P. Krichbaum, Anne Pagels,
      David Graham, Walter Alef, Arno Witzel and
      J. Anton Zensus: Max-Planck-Institut f\"ur Radioastronomie
      (MPIfR), Auf dem H\"ugel, 69, D-53121, Bonn, Germany
     (\email{iagudo, tkrichbaum, apagels, dgraham, walef,
      awitzel, azensus@mpifr-bonn.mpg.de}).
\item Uwe Bach: Istituto Nazionale di Astrofisica--INAF,
      Osservatorio Astronomico di Torino, via Osservatorio 20,
      10025 Pino Torinese--TO, Italy raiteri@to.astro.it
      (\email{bach@to.astro.it}).
\item Michael Bremer and Michael Grewing:
      Institut de Radio Astronomia Milim\'etrique
      (IRAM), 300 Rue de la Piscine, Domaine Universitaire
      de Grenoble, St. Martin d'H\`eres, F-38406, France
      (\email{bremer, greve, grewing@iram.fr}).
\item Harri Ter\"asranta: Mets\"ahovi radio observatory,
      Helsinki University of Technology, Mets\"ahovintie 114,
      02540 Kylm\"al\"a, Finland
      (\email{hte@kurp.hut.fi}).
}

\listofauthors{I. Agudo, T.~P. Krichbaum, U. Bach,
               A. Pagels, D. Graham, W. Alef, A. Witzel,
	       J.~A. Zensus, M. Bremer, M. Grewing
	       \& Ter\"asranta, H.}
\indexauthor{Agudo, I.}
\indexauthor{Krichbaum, T.~P.}
\indexauthor{Bach, U.}
\indexauthor{Pagels, A.}
\indexauthor{Graham, D.}
\indexauthor{Alef, W.}
\indexauthor{Witzel, A.}
\indexauthor{Zensus, J.~A.}
\indexauthor{Bremer, M.}
\indexauthor{Grewing, M.}
\indexauthor{Ter\"asranta, H.}

\abstract{The Global mm--VLBI Array is at present the most
          sensitive 3\,mm--VLBI interferometer and provides images
	  of up to 40 micro--arcsecond resolution.
	  Using this array, we have monitored the rotation of the innermost
	  jet in the quasar NRAO\,150, which shows an angular speed of
	  $\sim 7^{\circ}/\rm{yr}$.
	  Future 3\,mm arrays could include additional stations like
	  ALMA, GBT, LMT, CARMA, SRT, Yebes, Nobeyama and Noto, which
	  would allow to push VLBI at this wavelength to sensitivity
	  and image quality levels comparable to those of present VLBI
	  at centimeter wavelengths.
	  This would improve our knowledge of the accretion systems
	  and the magneto--hydrodynamics of the innermost jets in AGN
	  and microquasars.}

\resumen{El {\it Global mm--VLBI Array} es hoy d\'\i{}a el
          interfer\'ometro de VLBI a 3\,mm m\'as sensible y
	  proporciona im\'agenes con resoluciones angulares de
	  hasta 40 micro--segundos de arco.
	  Haciendo uso de este interfe\'ometro, hemos seguirdo
	  la evoluci\'on de la rotaci\'on en el plano del cielo
	  de la zona m\'as interna del jet en el qu\'asar NRAO\,150,
	  el cual presenta una velocidad angular de $\sim 7^{\circ}/$a\~{n}o.
	  Los futuros interfer\'ometros de 3\,mm podr\'\i{}an incluir
	  estaciones adicionales como ALMA, GBT, LMT, CARMA, SRT, Yebes,
          Nobeyama y Noto, que permitir\'\i{}an
	  forzar la t\'ecnica de VLBI a esta longitud de onda para
	  obtener niveles de sensibilidad y calidad de im\'agen
	  comparables a las de VLBI en longitudes de onda centim\'etricas.
	  Esto mejorar\'\i{}a
	  nuestro conocimiento de los sistemas de acrecimiento y la
	  magneto--hidrodin\'amica de las regiones m\'as internas
	  de los jets en AGN y microqu\'asares.}

\addkeyword{galaxies: active}
\addkeyword{galaxies: jets}
\addkeyword{quasars: individual (NRAO\,150)}
\addkeyword{radio continuum: general}
\addkeyword{techniques: high angular resolution}
\addkeyword{techniques: interferometric}

\begin{document}
\maketitle

\section{Introduction}
\label{sec:intr}

It has been shown in this conference that fundamental
questions related to the nature of the AGN are still open.
The accretion of material onto super--massive black
holes and the triggering of relativistic
jets (including their formation, acceleration and further
collimation) are some of the processes that still lack
a detailed understanding.
Observing with the highest angular resolution instruments
offers a good opportunity to learn more about these
processes through the study of the time evolution of
the jets.
An important effort has been made during the last decades
to bring the technique of millimeter Very Long Baseline
Interferometry (mm--VLBI) to progressively higher
sensitivities and shorter wavelengths, offering a
powerful tool to observe the innermost regions of
the jets and study the physics involved in their
behaviour.

During the last years, 7\,mm--VLBI observations, with angular
resolutions of up to $\sim 0.15$ milliarcseconds (mas), have addressed
the triggering of relativistic jets in AGN and their hydrodynamics.
Some particularly important results from these kind of observations
are the first size estimation of the radio--visible jet collimation
region ($\sim 1$\,pc from the core of the jet for
M\,87; Junor, Biretta \& Livio 1999) and the first measurement of
the distance from the central engine to the core of the jet
(of $\sim 0.3$\,pc for 3C\,120; Marscher et al. 2002).
Monitoring programs with adequate time sampling have also
allowed tests of relativistic hydrodynamic models in the
innermost regions of the jets in AGN (e.g. G\'omez et al. 2001
and Jorstad et al. 2005).

\begin{figure*}[t]
  \centering
  \includegraphics[width=14cm,clip]{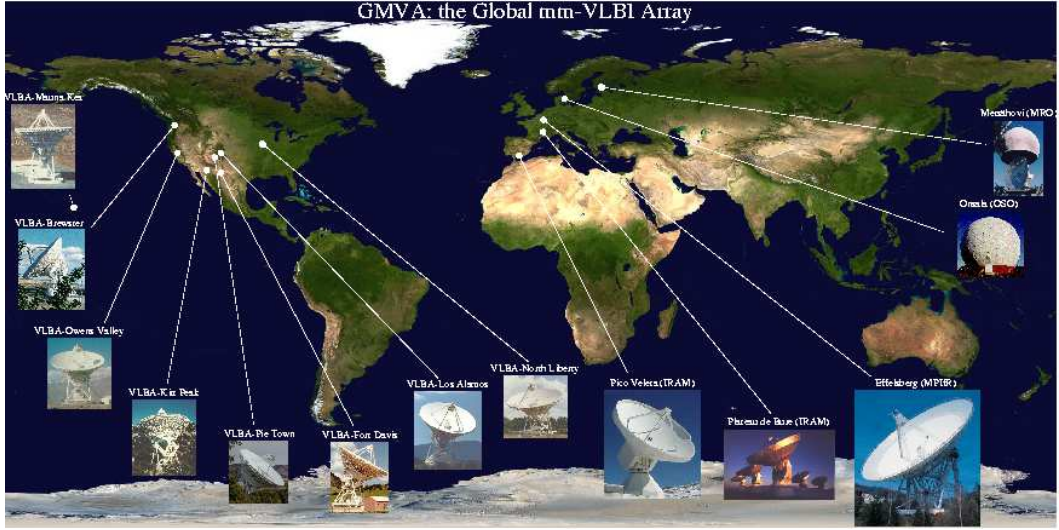}
  \caption{World distribution of the stations participating in the
          Global mm--VLBI Array.}
  \label{fig:GMVA}
\end{figure*}

At present, 3\,mm--VLBI offers an even better tool
to image deeper jet regions (i.e. closer than
$\sim 0.3$\,pc from the accretion system).
This is because of the lower jet opacities at this shorter
wavelength and to the larger resolving power at 3\,mm,
which can be up to three times higher than at 7\,mm.

\section{The GMVA: sensitive astronomy at
        40 micro--arcsecond resolution}
\label{sec:GMVA}

The most sensitive 3\,mm--VLBI instrument today is the
Global mm--VLBI Array
(GMVA\footnote{http://www.mpifr-bonn.mpg.de/div/vlbi/globalmm},
see Fig.~\ref{fig:GMVA}), composed of the Pico Veleta,
Plateau de Bure, Effelsberg, Onsala and Mets\"ahovi
stations, in addition to eight of the ten Very Long Baseline
Array (VLBA) antennas.
The GMVA achieves angular resolutions of up to
40\,$\mu$as with typical $7\sigma$ baseline sensitivities of
$80$--$100$\,mJy (adopting 20\,s coherence time, 100\,s
segmentation time and the standard GMVA recording rate
of 512\,Mbps).
This yields $7\sigma$ image sensitivities of $1$--$2$\,mJy/beam
(for 12\,h of observation and a duty cycle of 0.5).
With these characteristics the number of AGN which could
be imaged with high dynamic ranges ($\ge$100:1) is now
larger than 100.

In an attempt to obtain a deeper knowledge of the physics
in the innermost regions of jets in AGN, we have started
3\,mm--VLBI monitoring campaigns of some bright
sources. In this paper, we present recent
results about one of them: NRAO\,150.

\section{NRAO\,150: an unusual AGN  ``hidden" by the Milky Way}
\label{sec:NRAO150}

\begin{figure}[t]
  \includegraphics[width=\columnwidth,clip]{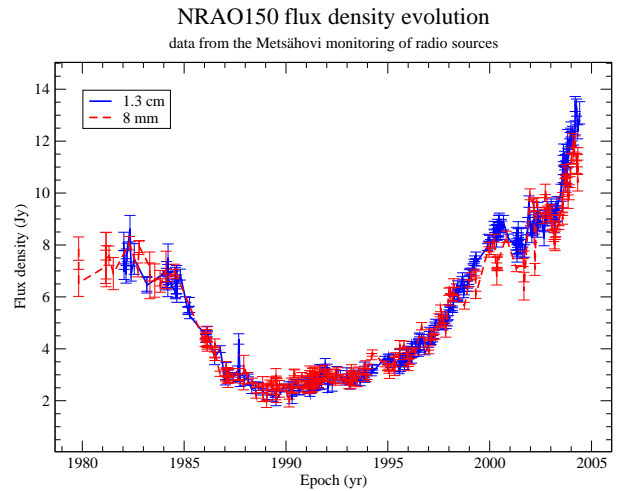}
  \caption{1.3\,cm and 8\,mm long term light curve of
          NRAO\,150 from 1987 to 2005. Data from the
	  Mets\"ahovi monitoring of radio sources.}
  \label{fig:metsa}
\end{figure}

\begin{figure*}[t]
  \centering
  \includegraphics[width=15cm,clip]{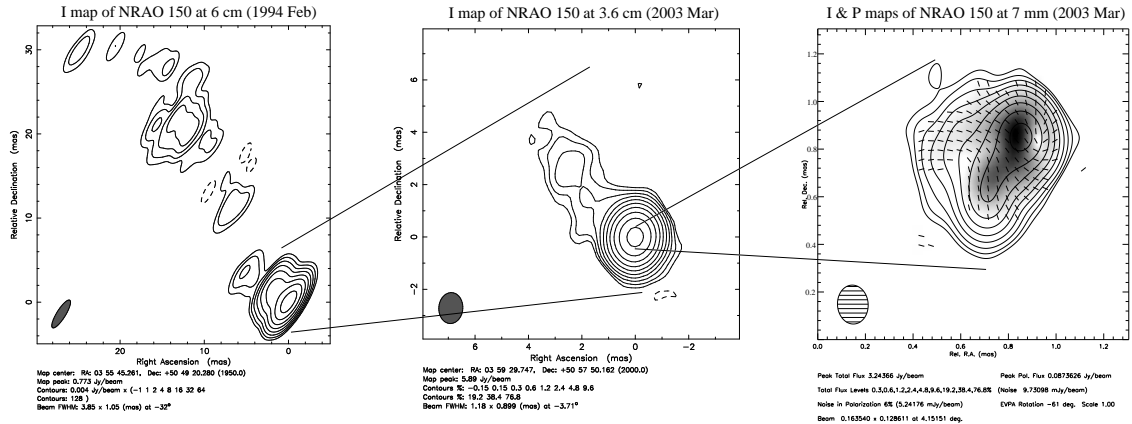}
  \caption{From left to right, 6\,cm, 3.6\,cm and 7\,mm VLBA images
          of the jet in NRAO\,150 obtained in February 1994
	  (for the 6\,cm map) and March 2003 (for those at 3.6\,cm and
	  7\,mm). The contours represent the observed total intensity.
	  For the 7\,mm map, the grey scale symbolizes the
	  polarized intensity and the superposed sticks the orientation
	  of the polarization electric vectors.
	  The higher resolution 7\,mm image shows a strong
	  misalignment between the cm and the mm jet of $\sim 120^{\circ}$.}
  \label{fig:misalign}
\end{figure*}

NRAO\,150 is an intense radio to mm source, which was first
cataloged by  Pauliny--Toth, Wade \& Heeschen
(1966).
The source has been monitored regularly in the radio and
millimeter bands since the beginning of the eighties.
Since then, its total flux density light curve has displayed
a quasi--sinusoidal behavior with a characteristic time--scale of
$\sim 20$--$25$)\,yr (see Ter\"asranta et al. 2004 and
Fig.~\ref{fig:metsa}).
The $1.3$\,cm light curve of the source peaked at
the beginning of 2004, when it displayed $\sim 11$\,Jy
(see Fig.~\ref{fig:metsa}).
NRAO\,150 lacks, up to now, an optical identification.
This is probably due to its low Galactic latitude
(-1.6$^{\circ}$), which causes strong Galactic extinction.
Although its distance is still unknown, we
hope to determine its redshift through an ongoing
spectroscopic project in the infrared band, at which
the source is not strongly absorbed.

On cm-VLBI scales, NRAO\,150 shows a compact core plus a
one--sided jet extending beyond 20 mas with a structural
position angle of $\sim 30^{\circ}$ (see Fig.~\ref{fig:misalign}).
Our new 7\,mm--VLBI observations, the first reported at this
wavelength, have revealed a strong misalignment,
of $\sim 120^{\circ}$, between the inner and outer jet
within its first 0.4\,mas (Fig.~\ref{fig:misalign}).

\subsection{The fastest jet rotation in an AGN at $\sim 7^{\circ}$/yr}
\label{sec:rot}

\begin{figure}[t]
  \centering
  \includegraphics[width=\columnwidth,clip]{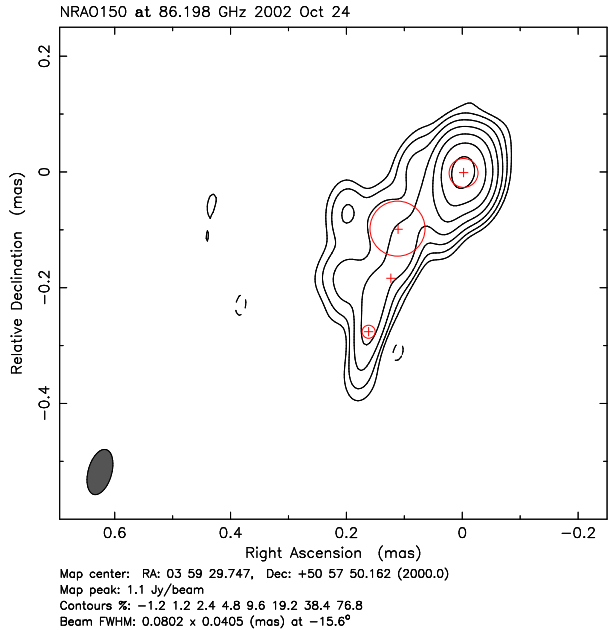}
  \caption{3\,mm--VLBI image of NRAO\,150 taken on October 2002.
           The positions of the fitted Gaussian components are
	   indicated by the crosses and the circles (of
	   radius equal to the FWHM of each Gaussian) symbolize
	   their size.}
  \label{fig:3mm}
\end{figure}

\begin{figure}[t]
  \centering
  \includegraphics[width=\columnwidth,clip]{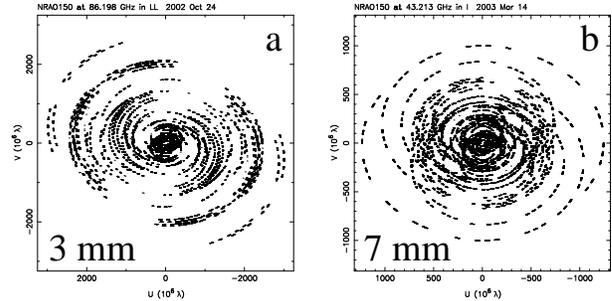}
  \caption{a) ({\it u},{\it v})--coverage for our 3\,mm--VLBI
           observation performed on 2002 October 24.
	   b) ({\it u},{\it v})--coverage for our 7\,mm--VLBI
           observation performed on 2003 March 14.}
  \label{fig:uv}
\end{figure}

Making use of the GMVA (and also of the former Coordinated
Millimeter VLBI Array, CMVA),
we have monitored the jet evolution since 1999 up to
date with observations performed about every six months.
Figure~\ref{fig:3mm} shows one of the resulting images from these
observations, which demonstrate the capability of the 3\,mm array to
probe the innermost jet structures with angular resolutions
of 40\,$\mu$as and dynamic ranges of $\sim$100:1.
The adequate image fidelity of our new 3\,mm observations
is also demonstrated by their ({\it u},{\it v})--coverage,
which is comparable to that of our 7\,mm observations
performed with the VLBA (see Fig.~\ref{fig:uv}).

\begin{figure}[t]
  \centering
  \includegraphics[width=\columnwidth,clip]{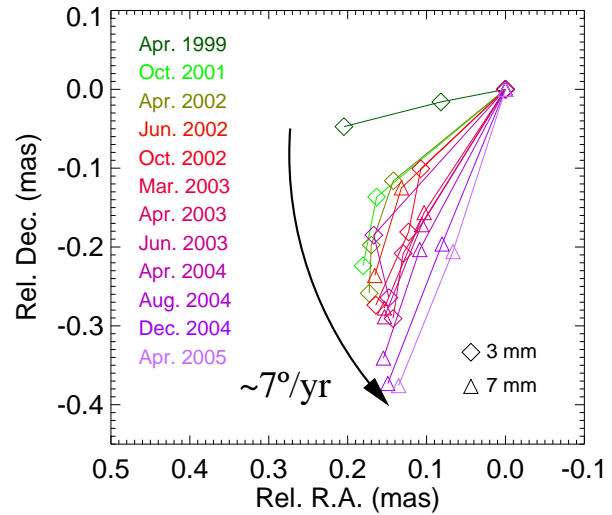}
  \caption{Position --with respect to the core of the jet emission--
           of the inner model--fit components
           in NRAO\,150.
	   Only results from observations performed between
	   1999 and 2005 at 3\,mm and 7\,mm are drawn.
	   The plot shows a fast change of the jet initial
	   direction with a mean angular speed of
	   $\sim 7^{\circ}$/year.}
  \label{fig:rot}
\end{figure}

The results from our new NRAO\,150 images have revealed a
clear angular rotation of the inner 0.4\,mas jet with a speed
of $\sim 7^{\circ}$/year -- projected on the plane of the sky --
(see Fig.~\ref{fig:rot}).
To our knowledge, this is so far the fastest jet rotation
reported for an AGN.

This phenomenon not only represents a likely explanation of
the large jet misalignment found in NRAO\,150, but it also
provides clues about the possible origin of the jet rotation.
It is reasonable to think that the quasi--sinusoidal light
curve of the source, its extreme jet misalignment and the
inner jet rotation are related.
In this case, a possible explanation of the NRAO\,150 evolution
would be a precession--like motion of the inner 0.4\,mas of the
jet.
This, together with projection effects and variable Doppler boosting
through small viewing angles, could explain the
strong jet misalignment, the jet rotation in the plane of the
sky and the $\sim 20$--$25$\,yr variability time--scale of the
radio light curves.
If, in the future, a significant correlation between these light
curves and the position angle of the inner jet is found, the
previous explanation will gain stronger support. In that case,
the possible period of the behavior of NRAO\,150 could be
measured from the light curves.

\section{Astrophysics from jet wobbling in AGN}
\label{sec:astro}

Like NRAO\,150, several other jets in AGN present
wobblings triggered in their innermost regions
(e.g., in BL\,Lac; Stirling et al. 2003 and Mutel \& Denn
2005, or in OJ\,287; Jorstad et al. 2005).
These wobblings can be induced by the development of helical
instabilities close to the jet base or by the precession of
the accretion disk.

For the former, jet--cloud interactions (G\'omez et al. 2000)
or dense ejections filling only part of the jet section could
be possible triggering perturbations.
However, they have not been extensively explored from the
theoretical point of view.
This is most likely due to our lack of knowledge of the jet
formation region and the lack of the adequate relativistic
magneto--hydrodynamic tools to study it.
Nonetheless, the subsequent development of Kelvin--Helmholtz
helical instabilities has been well studied (Hardee 2004
and references therein).

Disk precession seems to be nowadays the preferred
mechanism to test and model the quasi-regular jet structural
position and integrated emission variability of AGN.
Up to now, most precession models applied to AGN are
driven by a companion super--massive black hole or another
massive object (see e.g. Valtonen, Lehto \& Pietil\"a 1999,
for OJ\,287; Lister et al. 2003, for 4C~+12.50; Stirling
et al. 2003, for BL\,Lac; Caproni \& Abraham 2004
for 3C\,120; Lobanov \& Roland 2005, for 3C\,345 ).
However, alternative possibilities for accretion disk precession
-- and hence jet precession -- have appeared in the literature
during the last ten years (e.g., Schandl \& Meyer 1994;
Pringle 1996; Quillen 2001; Liu \& Melia 2002;
Lai 2003).
Among them, of special interest are the models from
Liu \& Melia (2002) and Lai (2003) which drive the precession
through intrinsic properties of the accretion system.
For that reason, they allow one to estimate or constrain
the possible black hole spin and accretion disk
density profile (for Sgr~A*, Liu \& Melia 2002; for a
set of eight AGN, Caproni, Mosquera--Cuesta \& Abraham
2004) and disk infall time (Lai 2003) from the observational
properties of the systems.

Although there is still no general paradigm to explain the
accretion disk (and jet) precession and other kinds of wobbling
for AGN, it is rather likely that, as they are triggered in the
innermost regions of the disks (and jets), their mechanisms have
to be tied to fundamental properties of these regions (i.e., close
to the accretion system).
Hence, further development of
models together with the appropriate characterization of the
observational properties of the innermost regions of jets in
AGN would place our understanding of the jet triggering region
and the super--massive accretion systems on firmer ground.

From the observational point of view, high resolution
mm-VLBI observations such as those presented here for NRAO\,150 
are of importance, as they allow to probe the innermost
(sub--pc scale) regions of jets in AGN.

\section{The future Global mm-VLBI Array}
\label{sec:futGMVA}

\subsection{Higher sensitivity, higher image fidelity
           and polarimetry}
\label{sec:futimprov}

Even with the good performance of the GMVA, it is desirable
to further improve the sensitivity and the quality of images.
This would increase the number of observable sources and
astrophysical scenarios.
The most direct way to achieve that is to increase the collecting
area of the present interferometric array.
For the near future, ALMA, the GBT, the LMT, CARMA, SRT, Yebes,
Nobeyama and Noto are some of the most sensitive stations
suitable to participate in 3\,mm--VLBI.
Together with them, the present GMVA would be able to achieve $7\sigma$
baseline sensitivities of (5 to 10)\,mJy, and $7\sigma$ image
sensitivities better than 0.1\,mJy/beam.
These estimates predict an \emph{increase, by a factor of 10}
with respect to the present GMVA sensitivity.
At the same time, the development of the VLBI technique
is providing ever faster data recording speeds.
For the next years, recording rates of at least 2\,Gbps are
expected (Garret 2003), which will increase the expected
sensitivities by an extra factor $\ge\sqrt{2}$.
Further improvements in coherence time for mm--VLBI, through
atmospheric phase correction methods (see Roy, Teuber \& Keller
2004 and also http://www.mpifr-bonn.mpg.de/staff/aroy/wvr.html),
are being developed at present.

\begin{figure}[t]
  \centering
  \includegraphics[bb= 104 111 492 732,width=\columnwidth,clip]{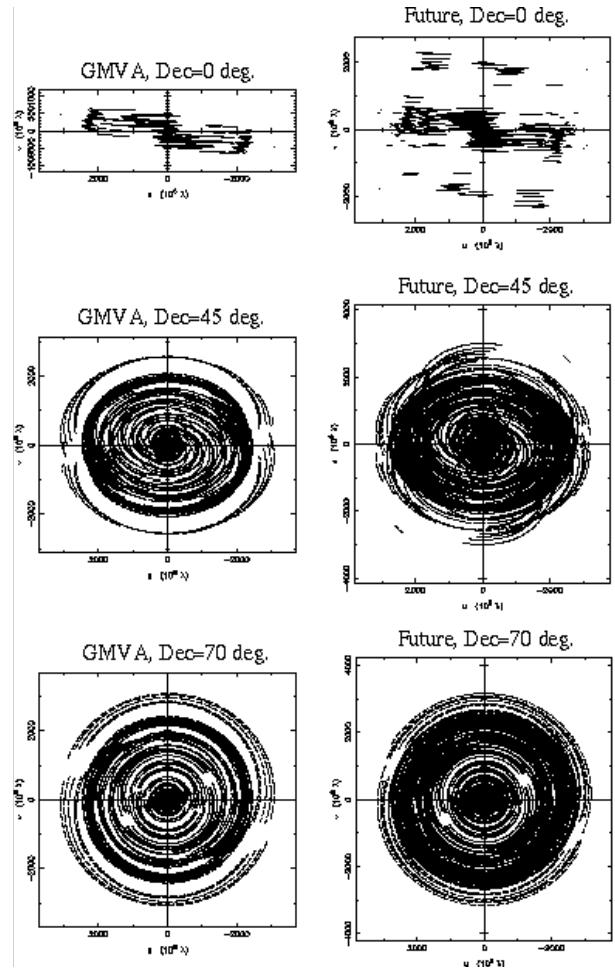}
  \caption{Simulations of the ({\it u},{\it v})--coverages
          of the present GMVA (left) and those of the GMVA
	  plus the suitable stations proposed in
	  \S~\ref{sec:futimprov} (right).
	  $0^{\circ}$, $45^{\circ}$ and $70^{\circ}$ of
	  source declinations are presented from top to
	  bottom.}
  \label{fig:uv_imp}
\end{figure}

But the proposed future array will not only produce an increase in
sensitivity. The new stations will also improve the
({\it u},{\it v})--coverage (see Fig.~\ref{fig:uv_imp}),
and so the image fidelity.
In addition, ALMA will improve the ({\it u},{\it v})--coverage
for sources with low declination (less than $20^{\circ}$) and
will facilitate the VLBI imaging of the Galactic Center source
Sgr\,A*.

Finally, high sensitivity 3mm--VLBI polarimetry is nowadays being
tested for the GMVA and it is expected to be offered as a standard
observing mode during the next years.

\subsection{Future science}
\label{sec:futscience}

If the proposed improvements are achieved in the future, images with
dynamic ranges of up to 1000:1 could be easily obtained.
This would place the sensitivity and image fidelity of 3\,mm--VLBI
at comparable levels than those of present cm-VLBI.
These achievements, together with
the possibility to obtain polarimetric images, would help to
(i) to study the triggering mechanisms of relativistic jets,
(ii) to probe their initial magnetic field configurations
(iii) and to better constrain the properties of their accreting
systems, for several hundreds or possibly thousands of jets in
AGN and microquasars.

\acknowledgements

I. Agudo and U. Bach acknowledge funding by the
European Commission through the TMR program HPRN-CT-2002-00321
(ENIGMA network).
We thank A.~L. Roy for helpful comments on this paper.
The VLBA is a facility of the National Radio Astronomy Observatory
of the USA, which is operated by Associated Universities, Inc., under
co-operative agreement with the National Science Foundation.


\begin{thebibliography}

\bibitem{} Caproni, A. \& Abraham, Z. 2004, MNRAS, 349, 1218

\bibitem{} Caproni, A., Mosquera--Cuesta, H. J. \& Abraham, Z. 2004,
           ApJ, 616, L99

\bibitem{} Garret, M.~A. 2003, ASP Conf. Serr, 306, 3

\bibitem{} G\'omez, J.~L., Marscher, A.~P., Alberdi, A., Jorstad, S.~G.
          \& Garc\'\i{}a--Mir\'o, C. 2000, Science, 289, 2317

\bibitem{} G\'omez, J.~L., Marscher, A.~P., Alberdi, A., Jorstad, S.~G.
          \& Agudo, I. 2001, ApJ, 561, L161


\bibitem{} Hardee, P.~E. 2004, Ap\&SS, 293, 117

\bibitem{} Jorstad, S.~G, Marscher, A.~P., Lister, M.~L. et al. 2005,
           AJ, 130, 1418

\bibitem{} Junor, W., Biretta, J.~A. \& Livio, M. 1999, Nature, 401, 891

\bibitem{} Lai, D. 2003, ApJ, 591, L119

\bibitem{} Lister, M.~L., Kellermann, K.~I., Vermeulen, R.~C., Cohen,
          M.~H., Zensus, J.~A. \& Ros, E. et al. 2003, ApJ, 584, 135

\bibitem{} Liu, S. \& Melia, F. 2002, ApJ, 573, L23

\bibitem{} Lobanov, A. P. \& Roland, J. 2005, A\&A, 431, 831

\bibitem{} Marscher, A.~P., Jorstad, S.~G., G\'omez, J.~L. et al. 2002,
           Nature, 417, 625

\bibitem{} Mutel, R. L. \& Denn, G. R. 2005, ApJ 623, 79

\bibitem{} Pauliny--Toth, I. I. K., Wade, C. M. \& Heeschen, D. S.
           1966, ApJS, 13,65

\bibitem{} Pringle, J.~E. 1996, MNRAS, 281, 857

\bibitem{} Quillen, A.~C. 2001, ApJ, 563, 313

\bibitem{} Roy, A. L., Teuber, U. \& Keller, R. 2004,
           in Proceedings of the 7th European VLBI Network
	   Symposium, ed. R. Bachiller, F. Colomer, J.~F.
	   Desmurls, P. de Vicente, 265


\bibitem{} Schandl, S., \& Meyer, F. 1994, A\&A, 289, 149

\bibitem{} Stirling, A.~M., Cawthorne, T.~M., Stevens, J.~A. et al.
           2003, MNRAS, 341, 405

\bibitem{} Ter\"asranta, H., Achren, J., Hanski, M. et al. 2004,
           A\&A, 427, 769

\bibitem{} Valtonen, M.~J., Lehto, H.~J. \& Pietil\"a, H. 1999,
           A\&A, 342, L29

\end{thebibliography}
\end{document}